\documentclass[authoryear, preprint]{elsarticle}

\usepackage[utf8]{inputenc}
\usepackage{amsmath, amsfonts, amssymb, amsthm} 
\usepackage{bm}                         
\usepackage{bbm}                        
\usepackage{graphicx, subcaption}        
\usepackage[english]{babel}             
\usepackage{natbib}                     
\usepackage{algpseudocode, algorithmicx, algorithm}   
\usepackage{geometry, setspace}         
\usepackage{hyperref,url}               
\usepackage{xcolor}                     
\usepackage{mathrsfs}
\usepackage{ifthen}
\usepackage{tikz}

\hypersetup{
  colorlinks=true,
  citecolor=orange,
  filecolor=red,
  linkcolor=blue,
  menucolor=red,
  runcolor=red,
  urlcolor=red
}

\theoremstyle{plain}

\newtheorem{Lemma}{Lemma}
\newtheorem*{Thm*}{Theorem}
\newtheorem*{Prop*}{Proposition}
\newtheorem*{Lemma*}{Lemma}
\theoremstyle{definition}

\newtheorem*{Defn*}{Definition}
\newtheorem*{Cond*}{Condition}
\theoremstyle{remark}

\newtheorem{Example}{Example}
\newtheorem*{Note*}{Note}
\newtheorem*{Example*}{Example}

\newcommand{\Th}{\textsuperscript{th}~}	

\newcommand{\ie}{\emph{i.e.\ }}

\newcommand{\place}{\mkern 2mu\cdot\mkern 2mu}
\newcommand{\infinity}{\infty}
\renewcommand{\to}{,\ldots,}

\def\app#1#2{%
  \mathrel{%
    \setbox0=\hbox{$#1\propto$}%
    \setbox2=\hbox{%
      \rlap{\hbox{$#1\sim$}}%
      \lower0.8\ht0\box0%
    }%
    \raise1\ht2\box2%
  }%
}

\DeclareMathSymbol{\minus}{\mathbin}{AMSa}{"39}

\renewcommand{\vector}[1]{\ensuremath{\left( #1 \right)}}
\newcommand{\set}[1]{\ensuremath{\left\{ #1 \right\}}}

\newcommand{\partfrac}[2]{\dfrac{\partial #1}{\partial #2}}

\newcommand{\fp}[3][]{\ensuremath{\operatorname{#2}\displaylimits_{#1}\!\left( #3 \right)}}
\newcommand{\fs}[3][]{\ensuremath{\operatorname{#2}\displaylimits_{#1}\!\left[ #3 \right]}}

\newcommand{\func}[3][]{\ensuremath{\fp[#1]{\operatorname{#2}}{#3}}}

\newcommand{\inlinefunc}[3][]{%
  \ensuremath{
    \ifthenelse{ \equal{#1}{} }
    {} {#1\!:\!}
    #2\!\mapsto\!#3
  }
}

\newcommand{\post}[1]{\ensuremath{\fp{\pi}{#1}}}
\newcommand{\prior}[1]{\ensuremath{\fp{\pi_0}{#1}}}
\newcommand{\prop}[1]{\ensuremath{\fp{q}{#1}}}

\newcommand{\logf}[1]{\ensuremath{\func{log}{#1}}}

\newcommand{\Gam}[1]{\ensuremath{\fp{\Gamma}{#1}}}

\newcommand{\grad}[2]{\ensuremath{\fp{\nabla #1}{#2}}}

\newcommand{\II}[1]{\fs{\mathbb{I}}{#1}} 
\newcommand{\PP}[1]{\fp{\mathbb{P}}{#1}}

\newcommand{\simiid}{\ensuremath{\overset{\textrm{iid}}{\sim}}}
\newcommand{\simind}{\ensuremath{\overset{\textrm{ind}}{\sim}}}

\newcommand{\Multinomial}[1]{\func{Multinomial}{#1}}
\newcommand{\Dirichlet}[1]{\func{Dirichlet}{#1}}

\newcommand{\Unif}[1]{\func{Unif}{#1}}
\newcommand{\Normal}[1]{\func{Normal}{#1}}

\newcommand{\Bernoulli}[1]{\func{Bernoulli}{#1}}
\newcommand{\Pois}[1]{\func{Pois}{#1}}

\newcommand{\bN}{\ensuremath{\bm{N}}}

\newcommand{\bZ}{\ensuremath{\bm{Z}}}

\newcommand{\bp}{\ensuremath{\bm{p}}}

\newcommand{\bw}{\ensuremath{\bm{w}}}

\newcommand{\bz}{\ensuremath{\bm{z}}}

\newcommand{\balpha}{\ensuremath{\bm{\alpha}}}

\newcommand{\btheta}{\ensuremath{\bm{\theta}}}

\newcommand{\bmu}{\ensuremath{\bm{\mu}}}

\newcommand{\brho}{\ensuremath{\bm{\rho}}}
\newcommand{\bsigma}{\ensuremath{\bm{\sigma}}}

\newcommand{\bvartheta}{\ensuremath{\bm{\vartheta}}}

\newcommand{\cA}{\ensuremath{\mathcal{A}}}

\newcommand{\cE}{\ensuremath{\mathcal{E}}}

\newcommand{\cG}{\ensuremath{\mathcal{G}}}

\newcommand{\cI}{\ensuremath{\mathcal{I}}}
\newcommand{\cJ}{\ensuremath{\mathcal{J}}}

\newcommand{\cS}{\ensuremath{\mathcal{S}}}

\newcommand{\cV}{\ensuremath{\mathcal{V}}}
\newcommand{\cW}{\ensuremath{\mathcal{W}}}

\definecolor{GreenYellow}       {RGB}{217, 229, 6}          
\definecolor{Yellow}            {RGB}{254, 223, 0}          
\definecolor{Goldenrod}         {RGB}{249, 214, 22}     
\definecolor{Dandelion}         {RGB}{253, 200, 47}     
\definecolor{Apricot}           {RGB}{255, 170, 123}    
\definecolor{Peach}             {RGB}{255, 127, 69}     
\definecolor{Melon}             {RGB}{255, 129, 141}    
\definecolor{YellowOrange}      {RGB}{240, 171, 0}          
\definecolor{Orange}            {RGB}{255, 88, 0}           
\definecolor{BurntOrange}       {RGB}{199, 98, 43}          
\definecolor{Bittersweet}       {RGB}{189, 79, 25}          
\definecolor{RedOrange}         {RGB}{222, 56, 49}          
\definecolor{Mahogany}          {RGB}{152, 50, 34}          
\definecolor{Maroon}            {RGB}{152, 30, 50}          
\definecolor{BrickRed}          {RGB}{170, 39, 47}          
\definecolor{Red}               {RGB}{255, 0, 0}        
\definecolor{BrilliantRed}      {RGB}{237, 41, 57}          
\definecolor{OrangeRed}         {RGB}{231, 58, 0}           
\definecolor{RubineRed}         {RGB}{202, 0, 93}       
\definecolor{WildStrawberry}    {RGB}{203, 0, 68}           
\definecolor{Salmon}            {RGB}{250, 147, 171}    
\definecolor{CarnationPink}     {RGB}{226, 110, 178}    
\definecolor{Magenta}           {RGB}{255, 0, 144}          
\definecolor{VioletRed}         {RGB}{215, 31, 133}     
\definecolor{Rhodamine}         {RGB}{224, 17, 157}     
\definecolor{Mulberry}          {RGB}{163, 26, 126}     
\definecolor{RedViolet}         {RGB}{161, 0, 107}          
\definecolor{Fuchsia}           {RGB}{155, 24, 137}     
\definecolor{Lavender}          {RGB}{240, 146, 205}    
\definecolor{Thistle}           {RGB}{222, 129, 211}    
\definecolor{Orchid}            {RGB}{201, 102, 205}    
\definecolor{DarkOrchid}        {RGB}{153, 50, 204}     
\definecolor{Purple}            {RGB}{182, 52, 187}     
\definecolor{Plum}              {RGB}{79, 50, 76}           
\definecolor{Violet}            {RGB}{75, 8, 161}           
\definecolor{RoyalPurple}       {RGB}{82, 35, 152}          
\definecolor{BlueViolet}        {RGB}{33, 7, 106}           
\definecolor{Periwinkle}        {RGB}{136, 132, 213}    
\definecolor{CadetBlue}         {RGB}{95, 158, 160}     
\definecolor{CornflowerBlue}    {RGB}{99, 177, 229}     
\definecolor{MidnightBlue}	{RGB}{0, 65, 101}           
\definecolor{NavyBlue}          {RGB}{0, 70, 173}       
\definecolor{RoyalBlue}         {RGB}{0, 35, 102}       
\definecolor{Blue}              {RGB}{0, 24, 168}       
\definecolor{Cerulean}          {RGB}{0, 122, 201}      
\definecolor{Cyan}              {RGB}{0, 159, 218}      
\definecolor{ProcessBlue}       {RGB}{0, 136, 206}      
\definecolor{SkyBlue}           {RGB}{91, 198, 232}     
\definecolor{Turquoise}         {RGB}{0, 255, 239}          
\definecolor{TealBlue}          {RGB}{0, 124, 146}          
\definecolor{Aquamarine}        {RGB}{0, 148, 179}          
\definecolor{BlueGreen}         {RGB}{0, 154, 166}          
\definecolor{Emerald}           {RGB}{80, 200, 120}     
\definecolor{JungleGreen}       {RGB}{0, 115, 99}           
\definecolor{SeaGreen}          {RGB}{0, 176, 146}          
\definecolor{Green}             {RGB}{0, 173, 131}          
\definecolor{ForestGreen}       {RGB}{0, 105, 60}           
\definecolor{PineGreen}         {RGB}{0, 98, 101}           
\definecolor{LimeGreen}         {RGB}{50, 205, 50}          
\definecolor{YellowGreen}       {RGB}{146, 212, 0}          
\definecolor{SpringGreen}       {RGB}{201, 221, 3}          
\definecolor{OliveGreen}        {RGB}{135, 136, 0}          
\definecolor{RawSienna}         {RGB}{149, 82, 20}          
\definecolor{Sepia}             {RGB}{98, 60, 27}           
\definecolor{Brown}             {RGB}{134, 67, 30}      
\definecolor{Tan}               {RGB}{210, 180, 140}	
\definecolor{Gray}              {RGB}{139, 141, 142}    
\definecolor{Black}{RGB}{30, 30, 30}       
\definecolor{White}{RGB}{255, 255, 255}    

\graphicspath{ {./paper/fig/} }

\journal{.}
\begin{document}

\begin{frontmatter}

  \title{Inference for a generalised stochastic block model with unknown
    number of blocks and non-conjugate edge models}

  \author[lancs]{Matthew Ludkin}
  \ead{m.ludkin1@lancaster.ac.uk}
  \address[lancs]{Mathematics and Statistics,
    Lancaster University,
    Lancaster,
    United Kingdom,
    LA1 4YW.
  }

  \begin{abstract}
    The stochastic block model (SBM) is a popular model for capturing
    community structure and interaction within a network.  Network
    data with non-Boolean edge weights is becoming commonplace;
    however, existing analysis methods convert such data to a binary
    representation to apply the SBM, leading to a loss of information.
    A generalisation of the SBM is considered, which allows
    edge weights to be modelled in their recorded state.  An effective
    reversible jump Markov chain Monte Carlo sampler is proposed for
    estimating the parameters and the number of blocks for this
    generalised SBM.  The methodology permits non-conjugate
    distributions for edge weights, which enable more flexible
    modelling than current methods as illustrated on synthetic data, a
    network of brain activity and an email communication network.
  \end{abstract}

\end{frontmatter}

\newlength{\subfigs}

\section{Introduction}
\label{sec:unb:intro}

Statistical analysis of networks has seen much growth in recent years
with the increasing availability of network data.  In this paper, a
network consists of a set of nodes, which can form pairwise
interactions. Each possible interaction is referred to as an \emph{edge},
with the value of that interaction called an \emph{edge weight}.

The aim of statistical network modelling is to describe the
edge weights with a probabilistic model, potentially performing
inference for model parameters.  Such models include the exponential
random graph \citep{snijders2006}, the class of latent space models
\citep{hoff2002} and the stochastic block model (SBM)
\citep{frank1982, holland1983}.  In the classic SBM, the set of nodes
is partitioned into \emph{blocks} such that the edge weight between two
nodes depends on their block memberships. There is a rich literature
on the SBM including both Bayesian and frequentist treatments.
Extensions to the SBM include restricting the SBM to only within-block
and between-block edge-weight distributions in the affiliation network
\citep{snijders1997, nowicki2001, copic2009}, multiple-block
memberships in the mixed-membership SBM \citep{airoldi2008},
degree-corrected SBM \citep{karrer2011}, and the infinite relational
model (IRM), \citep{kemp2006} where the number of blocks is treated as
unknown.  For a thorough review of the SBM and inference methods, see
\citet{matias2014}.

This paper considers two extensions to the SBM: (i) modelling general
edge weights (\ie non-binary interaction data) and (ii) estimating the
number of blocks.  Previous authors have attempted extension (i) with a weighted
or valued network \citep{jiang2009, mariadassou2010, ambroise2012} or
considering a time-series of edge weights \citep{matias2016, xin2017, ludkin2017}.
Multiple methods have been considered for extension (ii); these fall
into two main approaches: (a) a post-hoc analysis of multiple model
fits using model selection techniques, and (b) treating the number of
blocks as a random variable.  Approach (a) includes likelihood-based
methods using the Bayesian information criteria and its derivatives
\citep{daudin2008,latouche2012,wang2017,saldaa2017}, information-based
methods using minimum description lengths \citep{peixoto2013},
sequential testing by embedding successive block models with an
increasing number of blocks \citep{lei2016} and cross-validation
\citep{chen2016}. Approach (b) is achieved in a Bayesian framework by
setting a prior for the number of blocks. \citet{Geng2019} use a
mixture of finite mixtures representation, while the IRM
\citep{schmidt2013} uses a Chinese Restaurant Process (CRP)
\citep{gershman2012}.

Some authors \citep{morup2011,morup2012,schmidt2013,mcdaid2013} have
considered both extensions (i) and (ii) and posited collapsed Gibbs
samplers to perform inference on the number of blocks, node membership
and edge-weight model parameters. However, all of these methods require a
conjugate model for the edge-weight distributions.  This article aims
to achieve both extensions by generalising the SBM to arbitrary
edge-weight distributions and modelling the number of blocks in one
Bayesian framework \emph{without} the restriction of conjugate
edge-weight distributions.  This is highlighted in
Section~\ref{sec:unb:enron} where a negative binomial model is applied
to the edge weights within an email network.  Such a model cannot be
applied using existing methodology since no conjugate prior
distribution exists for the negative binomial with both parameters
unknown. This approach greatly broadens the applicability of the
general stochastic block model to network data with non-conjugate
edge-weight distributions.

The proposed methodology to perform inference is a Markov chain Monte
Carlo sampler which provides samples from the posterior distribution
of the block parameters, block memberships and number of blocks. The
sampling algorithm is inspired by \citet{green2001} -- a reversible
jump Markov chain Monte Carlo (RJMCMC) \citep{green1995} scheme using
split and merge proposals to explore the posterior by either combining
two blocks, or splitting a block into two. \citet{nobile2007,
  mcdaid2013} make use of a split-merge proposal, although due to the
conjugate models considered, they do not require parameter values.
The difficulty in designing an effective split-merge algorithm rests
on ensuring that parameter values are ``matched'' when changing
dimension.  Previous authors have proposed sampling algorithms, such
as the collapsed Gibbs sampler of \citet{mcdaid2013} -- for a given
node, the posterior probability of belonging to a given block is
computed with all other parameters fixed.  Under the collapsed regime,
 assigning a node to a new block is simple, since the
parameters have been integrated from the model.  In the case of
non-conjugate mixture models, the parameters are required to evaluate
the likelihood of such a re-assignment; this added complexity can be
handled within a full RJMCMC scheme as described in
Section~\ref{sec:unb:split-merge}.

The remainder of the paper is organised as follows: in
Section~\ref{sec:unb:model}, the specifics of the generalised SBM are
presented.  Section~\ref{sec:unb:split-merge} introduces the
split-merge sampling algorithm.  In Section~\ref{sec:unb:application},
the sampler is applied to simulated data, whilst in
Section~\ref{sec:unb:real-data}, the split-merge sampler is used to
analyse some real network data.  Finally, closing remarks and extensions
to the model and sampler are discussed in
Section~\ref{sec:unb:concluding-remarks}.

\section{A generalisation of the stochastic block model}
\label{sec:unb:model}
This section describes the stochastic block model and details the
generalisation to arbitrary edge-weight distributions for network
data.

Mathematically, a network is represented as a \emph{weighted graph}
$\cG = (\cV, \cE, \cW)$ where $\cV$ is the set of nodes,
$\cE \subseteq \cV\times \cV$ is the set of edges and $\cW$ is the set
of edge weights. This paper uses the shorthand
$ij \in \cE \implies (i,j) \in \cE$. The \emph{weight} of edge $ij$ is
denoted by $W_{ij} \in \cW$. To simplify exposition, it is assumed
that all edge weights are observed, \ie $\cE = \cV\times \cV$ and
$W_{ij} \in \cW$ for all $ij \in \cE$.  In this way, an un-weighted
graph $\cG=(\cV,\cE)$ can be viewed as a weighted graph
$\cG'=(\cV, \cE', \cW')$ with $\cE' = \cV\times\cV$, $W'_{ij} = 1$ if $ij \in \cE$ and $W_{ij}
= 0$ otherwise.
In the case where the network contains directed edges, the set $\cE$ consists of
ordered pairs such that $(i,j) \neq (j,i)$.

The canonical SBM \citep{holland1983,fienberg1985,wasserman1987} can
be viewed as such a weighted graph with $W_{ij}\in \set{0,1}$, a fixed
number of nodes ($|\cV| = N$) and $K$ blocks.  The nodes are
partitioned into blocks, with each node belonging to only one block.
Let $\bZ$ be the block indicator matrix with $Z_{ik} = 1$ if node $i$
belongs to block $k$ and 0 otherwise.  As such, $\bZ_i$ is a
one-of-$K$ indicator vector.  It is assumed that $\bZ_i$ is drawn from
a multinomial distribution with parameter $\brho$, a probability
vector of length $K$ which governs the block memberships.
The prior probability that a node belongs to block $k$ is given by
$\rho_k$.
Let $\btheta$ be a $K \times K$ matrix of edge-weight
parameters, such that $\vartheta_{kl}$ is the probability that
$W_{ij}=1$ between nodes $i$ and $j$ in blocks $k$ and $l$
respectively. Note $\vartheta_{kl} = \bZ_i^\top\bvartheta\bZ_j$. This
model is summarised in Equation~(\ref{eq:canonical-sbm}); first the
nodes are assigned to blocks, then -- given these block memberships --
the edge weights are drawn with parameters depending on the block
membership of the end nodes.

\begin{equation}
  \label{eq:canonical-sbm}
  \begin{aligned}
    \bZ_i | \brho &\simiid \Multinomial{\brho},\\
    W_{ij} | \bvartheta, \bZ &\simind \Bernoulli{\bZ_i'\bvartheta\bZ_j}.
  \end{aligned}
\end{equation}

In full generality, there are $K(K+1)/2$ free parameters in
$\bvartheta$ for an un-directed network (or $K^2$ for a directed network).  In
the affiliation model \citep{snijders1997, nowicki2001, copic2009},
$\bvartheta$ is restricted to two parameters, one each for
between-block ($\vartheta_{kl}, k \neq l$) and within-block
($\vartheta_{kk}$) interactions.

In this article, a parameterisation between these two extremes is
considered: let $\theta_k$ be the parameters governing edge weights
between nodes belonging to block $k$, and a global parameter
$\theta_0$ for edge weights between nodes in different blocks.  In
this way, the number of parameters is $K+1$, and grows linearly in the
number of blocks.  This model is appropriate for networks where
between-block connections are relatively homogeneous; for example, in
ecological contact networks, where herds of animals remain close
together for most of the time, with some interactions between herds.
Let $\btheta$ be the matrix of parameters with $\theta_{kk} =
\theta_k$ and $\theta_{kl} = \theta_0$ for $k=1 \to K, l\neq k$, then the
quadratic form $\bZ_i^\top\btheta\bZ_j$ picks the parameter governing
the edge weight $W_{ij}$.

With this parameterisation, the classic SBM in
Equation~(\ref{eq:canonical-sbm}) is extended to allow the number of
blocks to be random and to model general edge weights, such as count
or continuous data. Let $G$ and $G_0$ be the distribution on the
edges-weights and parameters respectively.  Prior parameters $\balpha$
are assigned to the block parameters $\btheta$.  Since the number of
blocks $K$ is considered unknown, a prior must be placed on both the
number of blocks and block memberships.  Let $F$ be a joint
distribution for $\vector{K, \bZ}$ with parameters $\gamma$ and
$\delta$ then the generalised form of the SBM considered in this paper
is:

\begin{equation}
  \label{eq:restricted-sbm}
  \begin{aligned}
    K, \bZ &\sim \fp{F}{\gamma, \delta},\\
    \theta_{k} &\simind \fp{G_0}{\balpha},\\
    W_{ij} | \btheta, \bZ &\simind \fp{G}{\bZ_i'\btheta\bZ_j }.
  \end{aligned}
\end{equation}

This framework may be extended to an edge-weight distribution $G$ with
multiple parameters.  For example, if $G$ represents the normal
distribution, then $\theta_k = \vector{\mu_k, \sigma_k}$ represents
the mean and standard deviation of the edge weights in block $k$.  In
this case, an additional subscript is required on $\theta_k$ such that
$\theta_{kp}$ is the $p$\Th parameter for block $k$.  In the normal
example, line 3 of Equation~(\ref{eq:restricted-sbm}) yields
$W_{ij} | \btheta, \bZ \simind \Normal{\bZ_i' \bmu \bZ_j,
  \bZ_i'\bsigma\bZ_j}$.

The choice of distributions for $G$ and $G_0$ is driven by the type
of edge weight considered (\ie edge weights representing counts could
be modelled using a Poisson distribution for $G$).
On the other hand, there is flexibility for
distribution $F$.  As discussed in \citet{Geng2019}, the popular
choice of the Chinese Restaurant Process (CRP) yields the undesirable
property that large probability is assigned to blocks with relatively
few nodes.  Indeed, \citet{Miller2018} show that using a CRP prior on
$(K, \bZ)$ in mixture models leads to inconsistent estimation of the
number of clusters, even in the asymptotic regime when $N$ tends to
infinity.  To circumvent this, \citet{Miller2018} propose using the
``mixture of finite mixtures approach'' (MFM) where the number of
blocks has an explicit prior distribution.  Let $F_0$ be a
distribution on $\set{1,2,3,\ldots}$ with parameter $\delta$,
then the prior for $(K, \bZ)$ considered in the remainder of the paper
is given in Equation~(\ref{eq:MFM-prior}):

\begin{equation}
  \label{eq:MFM-prior}
  \begin{aligned}
    K &\sim F_0(\delta),\\
    \rho |K &\simind \Dirichlet{\gamma, K},\\
    \bZ_i | \brho &\simind \Multinomial{\brho},
  \end{aligned}%
\end{equation}%
where $\Dirichlet{\gamma, K}$ is the symmetric Dirichlet distribution
on the $K-1$ simplex.
The \emph{size} of block $k$ is the number of nodes whose block
membership is $k$ and is given by $N_k = \sum_{i=1}^N
Z_{ik}$.
Let $\bN = \set{N_k : k = 1 \to K}$ be the set of block sizes, then the distribution for $\bN$ under the CRP and the MFM are:
\begin{align*}
  p_{\scriptscriptstyle{\text{CRP}}}(\bN) = \prod_{k=1}^K N_k^{-1}\quad\text{ vs. }\quad p_{\scriptscriptstyle{\text{MFM}}}(\bN) = \prod_{k=1}^K N_k^{\gamma-1}.
\end{align*}
Notice that the MFM gives comparatively less probability mass to small blocks than
the CRP.
Also, the distribution for the CRP is independent of $\gamma$.
Thus, the MFM approach gives more control over the prior block structure.

The parameter $\brho$ can be marginalised out of Equation~(\ref{eq:MFM-prior}) to obtain a prior
density for block memberships depending only on $K$ and $\gamma$ as such:
\begin{align*}
    \fp{f}{\bZ | \gamma, K} = \int_{\brho} \fp{f}{\bZ|\brho} \prior{\brho|\gamma} d\brho
    = \int_{\brho}\prod_{k=1}^K\rho_k^{N_{k} + \gamma+1} \frac{\Gam{K\gamma}}{\Gam{\gamma}^K} d\brho
    = \frac{\Gam{K\gamma}}{\Gam{\gamma}^K} \frac{\prod_{k=1}^K\Gam{\gamma + N_k}} {\Gam{K\gamma + N}},
\end{align*}
since $\sum_{k=1}^K N_k = N$ and where
$\Gam{a} = \int_0^\infinity x^{a-1}e^xdx$ is the gamma function; this
is referred to as the Dirichlet-Multinomial distribution.  Similarly,
the conditional distribution for the block membership of node $i$,
given $K$ and the other block memberships $\bZ_{-i}$ is:
\begin{align*}
    \fp{f}{\bZ_i | \bZ_{-i}, K, \gamma} &= \frac{\fp{f}{\bZ | \gamma, K}}{\fp{f}{\bZ_{-i} | \gamma, K}}
    = \frac{\prod_{k=1}^K\Gam{\gamma + N_k}} {\Gam{K\gamma + N}}\frac{\Gam{K\gamma + N - \sum_{k=1}^K Z_{ik}}}{\prod_{k=1}^K\Gam{\gamma + N_k- Z_{ik}}}\\
    &= \frac{1}{K\gamma + N - 1} \prod_{k=1}^K\frac{\Gam{\gamma + N_k}}{\Gam{\gamma + N_k-Z_{ik}}},
\end{align*}
since $\sum_{k=1}^K Z_{ik} = 1$ and $x\Gam{x} = \Gam{x+1}$. Therefore,
\begin{align*}
  \fp{f}{Z_{il}=1 | \bZ_{-i}, K, \gamma} &= \frac{\gamma + N_l - 1}{K\gamma + N - 1}.
\end{align*}

In the remainder of this article, the generalised SBM (GSBM) used is:
\begin{equation}
  \label{eq:general-sbm}
  \begin{aligned}
    K-1 &\sim \Pois{\delta},\\
    \bZ | K &\simind \fp{Dirichlet-Multinomial}{\gamma, K},\\
    \btheta_{k} &\simind \fp{G_0}{\balpha},\\
    W_{ij} | \btheta, \bZ &\simind \fp{G}{\bZ_i'\btheta\bZ_j },
  \end{aligned}
\end{equation}
where $G_0$ and $G$ are specified by the modeller.  The prior on
$(K, \bZ)$ will be referred to as the DMA($\gamma, \delta$)
(Dirichlet-Multinomial Allocation) prior.  When a model $G$ is
defined, we refer to the specific form of the model as $G$-SBM.

\section{Split-merge sampler}\label{sec:unb:split-merge}
This section discusses the benefit of split-merge steps
over Gibbs samplers for mixture models, describes the difficulty that arises
when designing split-merge moves for block membership in the GSBM, and
presents a split-merge RJMCMC sampler for the GSBM.
This algorithm draws samples from the posterior distribution of
$\vector{K, \bZ, \btheta}$.

For models containing a mixture component \citep[such as the block
structure in ][]{morup2012, mcdaid2013} a Gibbs sampler can get stuck
in local modes of the posterior.
Consider two ``true'' blocks $k$ and $l$ with sizes $N_k \geq N_l$ and
a state $s$ of a Gibbs sampler with a block $k^s$ consisting of all
nodes in true blocks $k$ and $l$.  For the Gibbs sampler to separate
the nodes in $k^s$ into blocks $k$ and $l$, it will require at least
$N_l$ steps, each of which takes a node assigned to $k^s$ and assigns
it to a new block $l^s$.  Each of these moves is quite unlikely,
especially if the parameters $\btheta_{k}, \btheta_l$ are close to
$\btheta_0$.  On the other hand, if all nodes could be moved at once,
then the proposal would be more likely to be accepted.  This is a
common problem with Gibbs sampling algorithms: the one-at-a-time
nature of the algorithm means large changes in posterior space are
unlikely, even if the combined changes increase the posterior
considerably.
One way to address this is to use a split-merge sampler.

Split-merge samplers have been developed for general mixture models
\citep{green2001}, with emphasis on a mixture of normal densities.  In
a standard parametric mixture model, each component has a different
form (either different distributions or different parameter values)
and each data point is drawn from a component of the mixture.  A
split-merge sampler applied to such a data set explores the possible
assignments of data points to components by successively proposing to
either merge two components together or split one component in two.
Care must be taken when designing such proposals: they must be an
isomorphism and differentiable to ensure the validity of the
underlying Markov chain.  Furthermore, to be efficient, a proposed
structure should have similar posterior support to the current
structure to give a reasonable  probability of acceptance.  Notice that,
since each data point belongs to one component, a split move which
assigns a data point to a new cluster will be penalised by the prior
on the number of components, but the likelihood will increase if the
parameter for the new component is a good fit for the assigned
data point.  Compare this to the latent block membership of the GSBM:
reassigning a node $i$ to a new block \emph{affects all nodes with an
  edge to $i$}. This implies that the prior will penalise the split
move for adding a block for the new node, and the likelihood will
penalise based on the $(N-1)$ edge weights incident to $i$.
Therefore, when considering split-merge samplers for the GSBM,
multiple edge weights are affected by changing the block membership of
one node; this fact complicates the design of a successful
proposal. 

The remainder of this section introduces the split-merge sampler for
the GSBM. The sampler consists of four moves: re-sampling parameter
values, splitting or merging blocks, reassigning nodes to the current
set of blocks, and adding or deleting an empty block.

Let $\vector{K^s, \bZ^s, \btheta^s}$ be the value of the parameters in
step $s$ of the sampler.  Values for parameter $\btheta$ given the
block structure can be sampled using any MCMC kernel.  In this
work, each $\theta_i$ is re-sampled using a random walk on a
transformed scale.  The difficult proposals are trans-dimensional:
merging and splitting blocks. These are described in the following
subsections.  The full split-merge algorithm is given in
Algorithm~\ref{alg:split-merge-sbm}.
\begin{algorithm}
  \begin{algorithmic}
    \State Inputs: edge-weight data $\bw$, prior parameters $\balpha, \gamma, \delta$, sampler parameters $\lambda, \nu, \sigma$.
    \State Draw $K^0, \bZ^0 \sim \fp{F_0}{\place | \gamma, \delta }$.
    \State Draw $\btheta^0 \sim \fp{G_0}{\place | \balpha }$.
    \For{$s=1 \to S$}
    \State Draw $\btheta^s \sim \func{Update}{\place | \bw, K^{s-1}, \bZ^{s-1}, \btheta^{s-1}, \balpha}$
    \State Let $K^s = K^{s-1}$
    \If{$K^s$=1}
    \State Propose a split
    \Else
    \State with probability 1/2 propose a split or a merge
    \EndIf
    \If{There are no empty blocks}
    \State Propose adding an empty block
    \Else
    \State with probability $\frac{N_\emptyset}{N_\emptyset+\nu}$ attempt deleting an empty block.
    \State or with probability $\frac{\nu}{N_\emptyset+\nu}$ attempt adding an empty block.
    \EndIf
    \For{$i=1\to N$}
    \For{$k=1\to K^s$}
    \State Let $p_k = \fp{g}{w_{i\place}| \bZ_{-i}, Z_{ik}=1, \btheta} \fp{f}{Z_{ik}=1 | \bZ_{-i}}$
    \EndFor
    \State Draw $\bZ'_i \sim \Multinomial{\bp}$
    \EndFor
    \State Store sample $\vector{\bZ^s, \btheta^s, K^s}$.
    \EndFor
    \State \Return samples $\bZ, \btheta, K$
  \end{algorithmic}
  \caption{Reversible jump Markov Chain Monte Carlo sampler for the
    GSBM with unknown $K$: split-merge algorithm.}
  \label{alg:split-merge-sbm}
\end{algorithm}

\subsection*{Merge move}
The merge proposal takes a state $(K^s, \bZ^s, \btheta^s)$ and
proposes a new state $(K', \bZ', \btheta')$. Such a move will reduce
the number of blocks by one: $K' = K^s -1$.  Firstly, two blocks $k$
and $l$ are sampled to merge -- possible mechanisms include choosing
blocks proportional to block size, inversely proportional to block
size, at random, etc.  In this paper, for simplicity, the pair $k,l$
is chosen with probability $1/K^s(K^s-1)$.  Secondly, the block
membership $\bZ'$ is updated. This is deterministic: any node that is
a member of block $k$ or $l$ in $\bZ^s$ is assigned to block $k'$ in
$\bZ'$. All other nodes keep their block assignment.  Next, the
parameter values are updated.  Following the recommendations of
\citet{green2001}, proposing a value $\btheta'_{k'}$ with similar
explanatory power as $\btheta_k$ and $\btheta_l$ should ensure that
$\btheta'_{k'}$ is well supported in the posterior.  A simple approach
is to take the mean value:
$\btheta'_{k'} = \btheta_k/2 + \btheta_l/2$; however, to allow more
flexibility in the sampler, an uneven merge is considered using a
weighted mean with tuning parameter $\lambda \in (0,1)$.  Since the
split move will invert the merge move, a \emph{matching function} $\operatorname{m}$
is required to ensure that parameters lie in the correct space. For
example, a rate parameter must be positive, whereby a suitable choice
for $\operatorname{m}$ is the exponential function.  Possible matching functions for
some common parameter spaces are shown in
Table~\ref{tab:possible-matching-funcs}.  The full parameter proposal
during a merge move is shown in Equation~(\ref{eq:merge-matched}):
\begin{equation}
  \fp{m}{\btheta'_{k'}} = \lambda \fp{m}{\btheta_{k}} + (1-\lambda) \fp{m}{\btheta_{l}} \label{eq:merge-matched}
\end{equation}
Finally, the acceptance probability $A_{merge}$ is computed (see
\ref{app:accept-prob-calc})
and the next state of the sampler
$(K^{s+1}, \bZ^{s+1}, \btheta^{s+1})$ is taken as
$(K', \bZ', \btheta')$ with probability $A_{merge}$, and as $(K^{s}, \bZ^{s}, \btheta^{s})$ otherwise.

\begin{table}[ht]
  \centering
  \caption{Possible matching functions to ensure parameters lie in the
    correct space.}
  \label{tab:possible-matching-funcs}
  \begin{tabular}[ht!]{ll}
    \hline
    Range for $\btheta$ & Possible matching function $\operatorname{m}$\\
    \hline
    $(\infinity,\infinity)$ & $\fp{m}{x} = x$\\
    $[0,\infinity)$ & $\fp{m}{x} = \logf{x}$\\
    $[0,1]$ & $\fp{m}{x} = \func{logit}{x} = \logf{x} - \logf{1-x}$\\
    \hline
  \end{tabular}
\end{table}

\subsection*{Split move}
The split proposal takes a state $(K^s, \bZ^s, \btheta^s)$ and
proposes a new state $(K', \bZ', \btheta')$ with $K' = K^s+1$.
Firstly, the block to split is chosen at random. Possible mechanisms
include sampling at random among the $K^s$ blocks,
proportional to block size, etc. In this paper the block is chosen
uniformly amongst the $K^s$ blocks.  To mirror the notation of the
merge move, the block to split is labelled $k'$, and the proposed new
blocks $k$ and $l$.

The first step in a split move determines the new block parameters.
This requires the inverse of Equation~(\ref{eq:merge-matched}).  On
top of this, an auxiliary variable $u'$ is needed to match the
dimension of the parameter space. In this work,
$u' \sim \Normal{0, \sigma^2}$ and represents the weighted difference
of the mapped parameters $\fp{m}{\theta_k}$ and $\fp{m}{\theta_l}$. The
parameter split is thus:
\begin{align*}
  \fp{m}{\btheta_{k}} &= \frac{\fp{m}{\btheta'_{k'}} + u'}{2\lambda'}
  \fp{m}{\btheta_{l}} &= \frac{\fp{m}{\btheta'_{k'}} - u'}{2(1-\lambda')}
\end{align*}

Note that the dimension-matching criterion of RJMCMC \citep{green1995}
is achieved since the vectors $\vector{\btheta'_{k'}, u', \lambda'}$ and
$\vector{\btheta_{k}, \btheta_{l}, \lambda}$ have the same cardinality.

To determine $\bZ'$, the nodes assigned to block $k'$ in $\bZ^s$ are
reassigned to blocks $k$ and $l$. In a similar fashion to \citet{green2001}, nodes are assigned
sequentially to either block $k$ or $l$ proportional to the
model likelihood.
It is not possible to compute the full likelihood
during this procedure for the GSBM because edge weights exist
between all nodes.
Specifically, let $i$ and $j$ be the only nodes in block $k'$.
Choosing to assign $i$ to block $k$ or $l$ proportional to the
likelihood requires knowledge of the block membership of $j$, which
does not yet exist.
The quantity can be calculated in principle by looking at all the
possible allocations of the nodes in block $k$ to $k'$ and $l'$.
This operation is expensive; instead, it is estimated by the following
sequential process:

First, all nodes in block $k'$ are unassigned and placed in a holding
set $\cI$. The set of remaining nodes is labelled $\cJ$ and the
current set of block assignments $\bZ_{\cJ}$.
Take a permutation $\sigma(\cI)$ of $\cI$ -- this is the order in
which nodes will be reassigned to block $k$ or $l$.

When assigning node $i$, the following quantity can be calculated:
\begin{equation*}
  \label{eq:rj-sequential-probability}
  \prop{Z'_i = k'} = \frac{\fp{f}{\bw | Z'_i = k', \bZ'_{\cJ}, \btheta'}} {\fp{f}{\bw | Z'_i = k', \bZ'_{\cJ}, \btheta'} + \fp{f}{\bw | Z'_i = l', \bZ'_{\cJ}, \btheta'}}.
\end{equation*}
Node $i$ is then assigned to block $k$ with probability $\prop{Z'_i =
  k}$ and to block $l$ otherwise.
Once assigned, $i$ is moved from $\cI$ to $\cJ$ for the next assignment.

The total proposal probability of the new block assignment is thus:
\begin{equation*}
  \label{eq:rj-permutation-probability}
  \prop{\bZ'} = \prod_{i \in \sigma(\cI)} \prop{Z'_i = k}^{\II{Z'_i = k'}}(1-\prop{Z'_i = k})^{\II{Z'_i = l'}}.
\end{equation*}

Finally, the proposed split is accepted as the next state of the
sampler with probability $A_{split}$ as in
Equation~(\ref{eq:rj-a-split-full}), Appendix A.

\subsection*{Gibbs reassignment move}
To allow the sampler to explore the parameter space, an additional two
moves are included: a Gibbs-like move (which allocates each node to a
block proportional to the posterior density) and a move that allows
the addition and deletion of empty blocks.

The Gibbs-like allocation move for node $i$ computes the conditional
posterior value for $i$ being a member of each of the $K$ blocks in
the current state of the sampler. Since $K$ is finite, this set of
posterior values can trivially be normalised to a probability vector,
such that $p_{ik}$ is the probability that node $i$ is reassigned to
block $k$.  Thanks to the structure of the GSBM, $p_{ik}$ can be
written as the product of two densities: the posterior density of edge
weights to nodes in block $k$, and the posterior density of edge
weights to nodes in other blocks:

\begin{equation*}
  \label{eq:split-merge-gibbs-alloc}
  \begin{aligned}
    p_{ik} &= \fp{p}{Z_{ik} = 1 | \bZ_{-i}, \bw, \btheta}, \\
    &\propto \fp{f}{\bZ_{ik}=1|\bZ_{-i}} \prod_{j \neq i} \fp{g}{w_{ij} | \bZ_j, bZ_{ik}=1, \btheta},\\
    &= \fp{f}{\bZ_{ik}=1|\bZ_{-i}}  \prod_{j \neq i} \fp{g}{w_{ij} | \btheta_k}^{\bZ_{jk}}  \fp{g}{w_{ij} | \btheta_0}^{1-\bZ_{jk}}.
  \end{aligned}
\end{equation*}

Notice it is possible to reassign $i$ to its current block.
This move, as well as the split move, can leave a block empty;
waiting for the sampler to merge an empty block with another block can
leave empty blocks in the sampler state for some time, adding to the
uncertainty around the number of blocks $K$. A proposal that addresses
these concerns is considered in the next section.

\subsection*{Add or delete empty blocks}
The second extension allows for the deletion and addition of empty
blocks; the \emph{delete empty block} move is the inverse of \emph{add
  empty block}.  During the \emph{delete empty block} move, a
candidate block is chosen at random from the current set of empty
blocks.  When an empty block is added, it is given the label $K+1$.
For simplicity, when an add/delete move is attempted, the probability
of adding a block is chosen proportional to a sampler parameter $\nu$.
The probability of choosing to delete an empty block is proportional
to the number of empty blocks in the current state, $N_\emptyset$.
Note that the likelihood of the edge weights does not change with the
addition of empty blocks since the entire node structure remains
unaffected.  When a block is added, a parameter $\btheta^*$ is drawn
from the prior distribution $G_0$.  The acceptance probabilities of
the add and delete empty block moves are calculated as:

\begin{equation*}
  A_{add} = \frac{\prior{K+1, \bZ}}{\prior{K, \bZ}}
  \frac{\nu + N_\emptyset}{\nu(\nu + N_\emptyset+1)}, \quad \text{ and } \quad A_{del} = \frac{\prior{K-1, \bZ}}{\prior{K, \bZ}}
  \frac{\nu(\nu + N_\emptyset)}{\nu + N_\emptyset-1}.
\end{equation*}

The sampler is implemented in the \texttt{R} package ``SBMSplitMerge''
\citet{CRANSBMSplitMerge}. This package is used to perform the
inference in the following sections.

\section{Simulated data}\label{sec:unb:application}
In this section, the split-merge sampler of
Section~\ref{sec:unb:split-merge} is demonstrated on simulated
data. The scripts to generate these example networks, run the sampler, and produce the
figures (as well as the data in Section~\ref{sec:unb:real-data}) are
available on GitHub
(\url{https://github.com/ludkinm/SBMSplitMerge/releases/tag/CRAN-1.1.1}).

\begin{figure}[ht!]
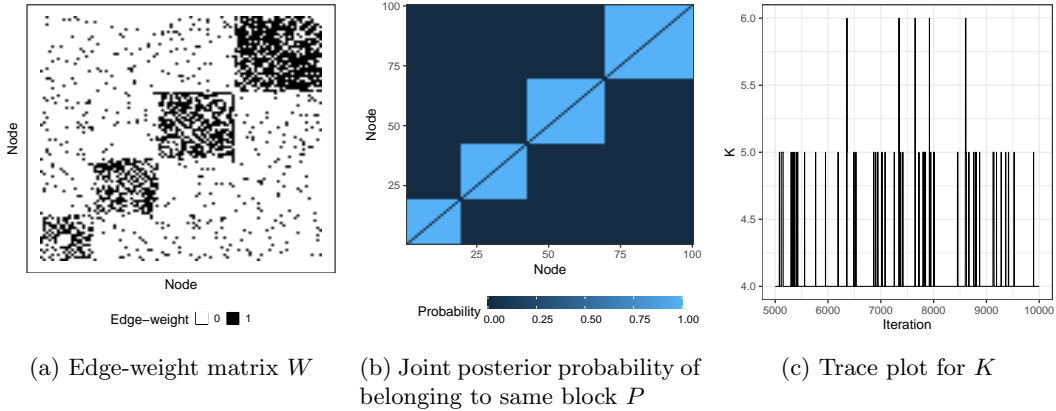

  \setlength{\subfigs}{0.3\textwidth}
  \centering
  \begin{subfigure}[t]{\subfigs}
    \centering
    \includegraphics[width=\textwidth]{{{bern/rj_edges}}}
    \caption{Edge-weight matrix $W$}
    \label{fig:bern-rj-edges}
  \end{subfigure}%
  ~
  \begin{subfigure}[t]{\subfigs}
    \centering
    \includegraphics[width=\textwidth]{{{bern/rj_post_pairs_10000}}}
    \caption{Joint posterior probability of belonging to same block $P$}
    \label{fig:bern-rj-pp}
  \end{subfigure}%
  ~
  \begin{subfigure}[t]{\subfigs}
    \centering
    \includegraphics[width=\textwidth]{{{bern/rj_num_blocks_trace_10000}}}
    \caption{Trace plot for $K$}
    \label{fig:bern-rj-nb}
  \end{subfigure}%
  \caption{Bernoulli edge weights: adjacency matrix and posterior summaries for block
    membership and number of blocks $K$.}
  \label{fig:bern-post}
\end{figure}

\begin{figure}[ht!]
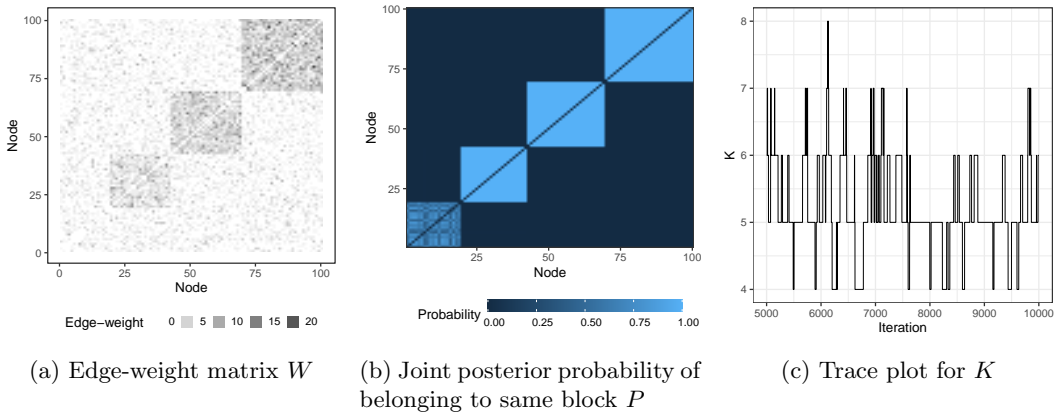

  \setlength{\subfigs}{0.3\textwidth}
  \centering
  \begin{subfigure}[t]{\subfigs}
    \centering
    \includegraphics[width=\textwidth]{{{nbin/rj_edges}}}
    \caption{Edge-weight matrix $W$}
    \label{fig:nbin-rj-edges}
  \end{subfigure}%
~
  \begin{subfigure}[t]{\subfigs}
    \centering
    \includegraphics[width=\textwidth]{{{nbin/rj_post_pairs_10000}}}
    \caption{Joint posterior probability of belonging to same block $P$}
    \label{fig:nbin-rj-pp}
  \end{subfigure}%
  ~
  \begin{subfigure}[t]{\subfigs}
    \centering
    \includegraphics[width=\textwidth]{{{nbin/rj_num_blocks_trace_10000}}}
    \caption{Trace plot for $K$}
    \label{fig:nbin-rj-nb}
  \end{subfigure}%
  \caption{Negative binomial edge weights: adjacency matrix and posterior summaries for block
    membership and number of blocks $K$.}
  \label{fig:nbin-post}
\end{figure}

Two data sets are considered. Both consist of 100 nodes split into
four blocks with sizes 19, 23, 27 and 31. Each network has the same
block structure.  The first data set uses a Bernoulli distribution as
its edge-weight distribution $G$.  The second data set uses a
generalised negative binomial distribution.
Data was simulated from the edge-weight distributions with and plotted in
Figure~\ref{fig:bern-rj-edges} for the Bernoulli data set, then
Figure~\ref{fig:nbin-rj-edges} for the negative binomial.

The generalised negative
binomial distribution is parameterised by the real-valued ``number of
failures'' $r > 0$ and success probability $p\in [0,1]$.  If
$X \sim \func{NegBin}{r,p}$ then:
\[
  \PP{X=x} = \frac{\Gam{x+r}}{\Gam{r} x!} p^r (1-p)^x, \text{ for } x=0,1,2,\ldots
\]
Notice that the Bernoulli distribution admits a conjugate prior;
therefore, existing samplers, such as those introduced by
\citet{morup2012} and \citet{mcdaid2013}, could be applied.  However,
for the negative binomial with both $r$ and $p$ unknown, no conjugate
prior exists.

To apply the GSBM, the prior on $K$ and $\bZ$ was set to a DMA
distribution with hyperparameters set to $(\gamma, \delta) = (1, 10)$.
The parameter values used for each of the edge-weight models is given
in Table~\ref{tab:sim-params}.  For the network with
Bernoulli-distributed edge weights, the uniform prior Beta(1, 1) was
applied to each parameter $\btheta$.  In the negative binomial network
with both parameters unknown, a Beta(1, 1) distribution is placed on
the probability parameter $p$ and the prior for $r$ is set to Gamma(1,
1).

\begin{table}[ht]
  \centering
  \caption{Simulated data parameter values for each edge-weight
    distribution.}
  \label{tab:sim-params}
  \begin{tabular}[ht]{lrrrrr}
    \hline
    Parameter & $\theta_0$ & $\theta_1$ & $\theta_2$ & $\theta_3$ & $\theta_4$\\\hline
    Bernoulli($p$) &  0.05 & 0.4 & 0.5 & 0.6 & 0.7 \\
    Negative binomial($p,r$) &  (0.5, 1) & (0.5, 1) & (0.5, 4) & (0.5, 5) & (0.5, 6) \\
    \hline
  \end{tabular}
\end{table}

In both cases, a random walk Metropolis-Hastings step was applied to
$\btheta$ on a transformed scale with standard-deviation 0.1.  A draw
from the prior was taken as the initial state then the split-merge
sampler of Section~\ref{sec:unb:split-merge} ran for 10,000 iterations
with 5000 iterations discarded as burn-in.

To evaluate the performance of the algorithm, the ability to detect
the true number of blocks, block structure and parameter values are
considered.  To measure the ability to detect block structure, the
posterior joint probabilities that two nodes belong to the same block
are calculated after burn-in, via:
\begin{equation}
  \label{eq:post_pairs}
  P_{ij} = \frac{1}{| \cS |}\sum_{s\in \cS} \II{Z_{is} = Z_{js}},
\end{equation}
where $\cS$ contains the indices of samples remaining after burn-in.

The parameter estimates can be compared to the true values in
Table~\ref{tab:sim-params}.  Note that the model in
Equation~(\ref{eq:general-sbm}) is invariant to a permutation of the
block labels; this implies that the true and inferred structure may be
the same up to a permutation of the block labels. To correct for this
phenomenon, a permutation of the modal block labels under the MCMC to
the true labels is derived and applied to the parameters and block
labels in the Markov chain (Details are given in
\ref{app:matching}).  Note this matching is only required to
compare the true parameter values to the MCMC output.

The posterior joint probability that two nodes are in the same block
(after burn-in) is displayed for the Bernoulli network in
Figure~\ref{fig:bern-rj-pp}. This matches the truth very well: nodes
who truly are in the same block have high posterior probability of
being assigned to the same block (Equation~\ref{eq:post_pairs}), and
nodes who are not in the same block have low posterior probability.
The trace plot for $K$ shows that for most iterations the sampler had
four blocks, matching the truth, but explored some states with five or
six blocks.  The posterior modes of the parameters, and the 5\% and
95\% posterior confidence intervals are shown in
Table~\ref{tab:est-params}.  The posterior modes are all close to the
true values in Table~\ref{tab:sim-params} for the Bernoulli network.

For the negative binomial network, Figure~\ref{fig:nbin-rj-pp} shows
that blocks 2, 3 and 4 are well identified by the sampler.  As for
the block 1, recall $\theta_0=\theta_1$ in the true parameters; this
gives no structure to block 1.  Indeed, one could reassign the nodes
in block 1 arbitrarily between two blocks 1a and 1b with
$\theta_{1a} = \theta_{1b} = \theta_1$ and the likelihood would be
unchanged. (Note this is not true for block $k=2,3,4$ since some
within-block interactions governed by $\theta_k \gg \theta_0$ would be
governed by $\theta_0$ under such a reassignment.) The sampler is able
to explore regions of the posterior where nodes in block 1 are
separate from the other nodes, as seen by the low probability region
in the off-diagonal in Figure~\ref{fig:nbin-rj-pp}. There is
uncertainty around if the nodes in block 1 are in the same block as
indicated by the range of posterior probabilities in the lower left
block of Figure~\ref{fig:nbin-rj-pp}.  The estimated parameter values
in Table~\ref{tab:est-params} lead to similar conclusions: the
estimates for parameters $\theta_0, \theta_2, \theta_3$ and $\theta_4$
are good, but, the poor specification of block 1 leads to poor
estimates of $\theta_1$.

\begin{table}[ht]
  \centering
  \caption{Mode, 5\% and 95\% posterior quantiles for parameters in
    example networks.}
  \label{tab:est-params}
  \begin{tabular}[ht]{lrrr}
    \hline
    Model & Bernoulli & Negative Binomial & Negative Binomial\\
    Parameter & $p$ & $p$ & $r$\\
    \hline
    $\theta_0$ & 0.052 (0.046, 0.058) & 0.472 (0.442, 0.497) & 0.895 (0.801, 0.978)\\
    $\theta_1$ & 0.425 (0.366, 0.491) & 0.436 (0.059, 0.997) & 0.642 (0.001, 1.575)\\
    $\theta_2$ & 0.506 (0.453, 0.557) & 0.467 (0.392, 0.536) & 3.196 (2.410, 4.126)\\
    $\theta_3$ & 0.638 (0.598, 0.677) & 0.536 (0.472, 0.600) & 5.545 (4.330, 7.183)\\
    $\theta_4$ & 0.678 (0.643, 0.714) & 0.477 (0.425, 0.532) & 5.392 (4.480, 6.692)\\
    \hline
  \end{tabular}
\end{table}

Assessing the convergence of a reversible jump Markov chain is
non-trivial.  Two techniques are applied in this section: (i) applying
the Gelman-Rubin convergence statistic \citep{gelman1992} to a summary
statistic and (ii) starting two independent samplers from extreme
block configurations -- one with all nodes assigned to one block and
the other with each node assigned to a unique block.

In the first case, the mean and variance of the parameter values are
used as summary statistics of the sampler performance, which are
recorded at every iteration of the sampler.  The
Gelman-Rubin statistics for the sampler for each
model are shown in Table~\ref{tab:rubin-gelman} based on 30
independent chains.  These values are close to 1, indicating
that convergence appears to have occurred during the first 10,000
iterations.

\begin{table}[ht!]
  \centering
  \begin{tabular}[ht!]{lrrrrr}
    \hline
    Model          &       Bernoulli & Negative binomial\\
    \hline
    Mean           & 1.0005 (1.0007) & 1.0098 (1.0153)\\
    \hline
    Variance       & 1.0005 (1.0006) & 1.0069 (1.0106)\\
    \hline
  \end{tabular}
  \caption{Rubin-Gelman statistics (and upper bound of 95\% confidence interval) for each model with 30 independent chains of 10000 iterations.}
  \label{tab:rubin-gelman}
\end{table}

\begin{figure}[ht!]
  \setlength{\subfigs}{0.45\textwidth}
  \centering
  \begin{subfigure}[t]{\subfigs}
    \centering
    \includegraphics[width=\textwidth]{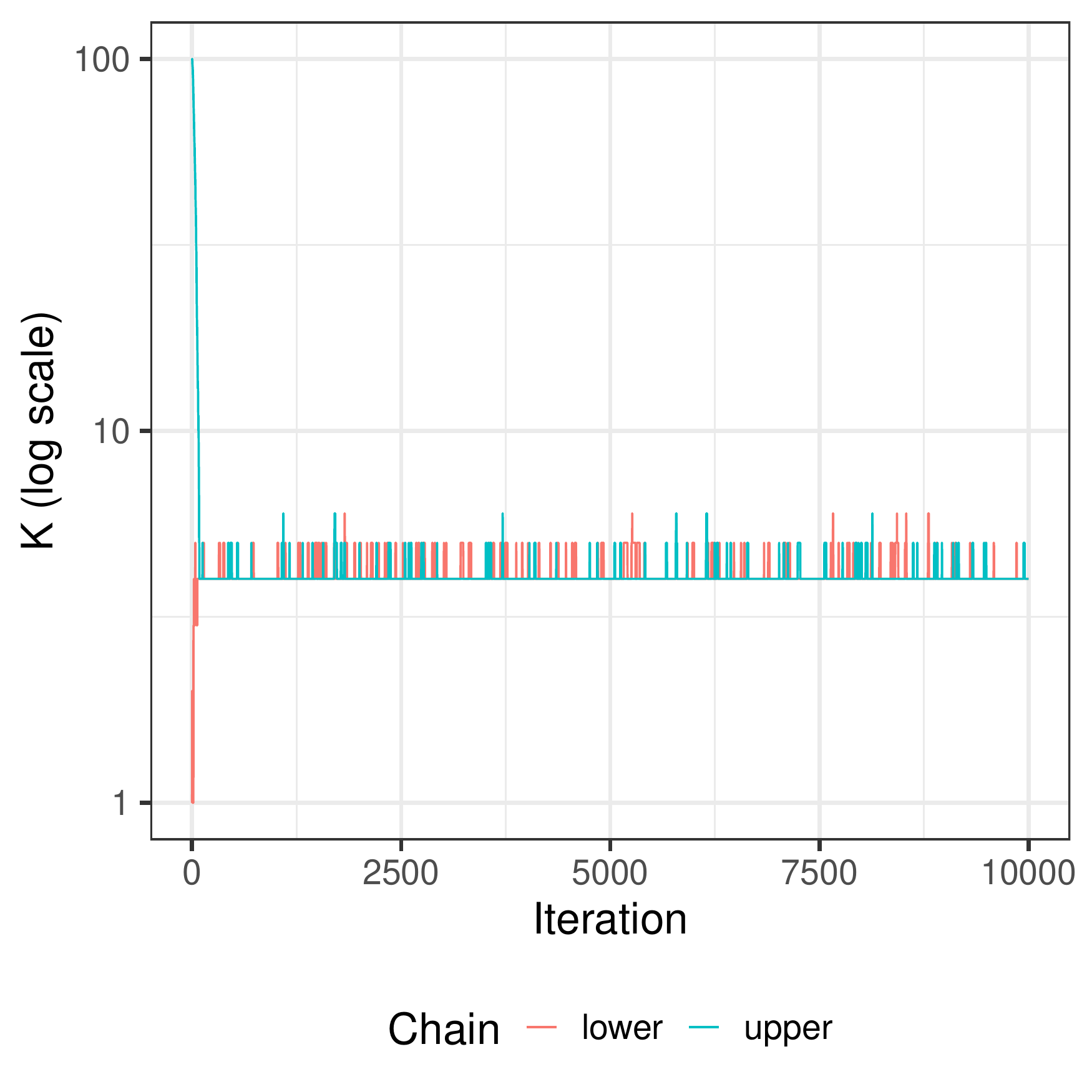}
    \caption{Bernoulli: Perfect simulation trace plot for $K$}
    \label{fig:bern-convergence-rj}
  \end{subfigure}
  ~
  \begin{subfigure}[t]{\subfigs}
    \centering
    \includegraphics[width=\textwidth]{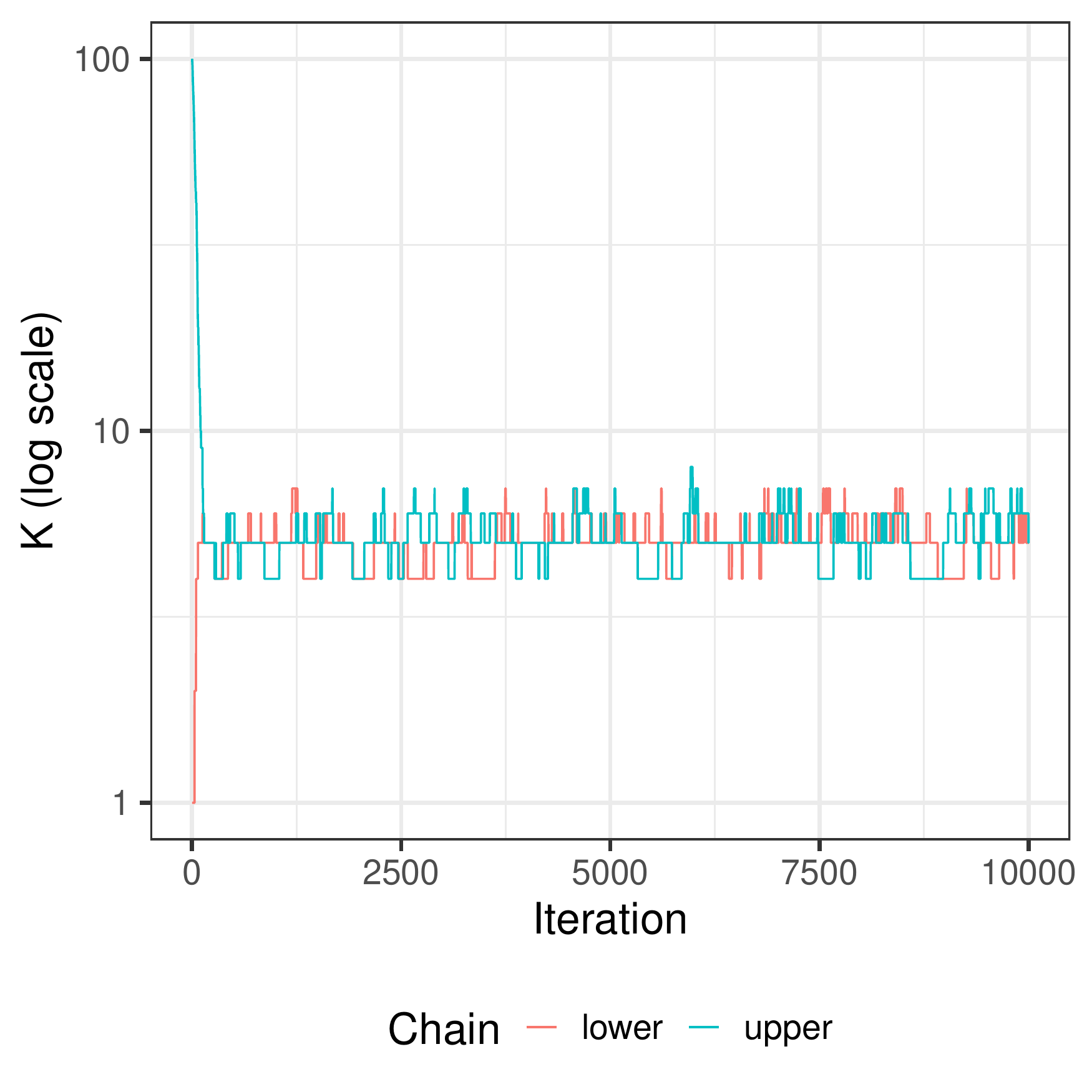}
    \caption{Negative binomial: Perfect simulation trace plot for $K$}
    \label{fig:nbin-convergence-rj}
  \end{subfigure}%
  \caption{Trace plots for number of blocks $K$ in example networks.
    Two chains are simulated in each case: the ``lower chain'' with all nodes initially in one block (orange line) and the ``upper chain'' with all nodes initially assigned to different blocks (teal line).}
  \label{fig:convergence-bounding}
\end{figure}

The second technique for assessing convergence is inspired by perfect
simulation: starting two samplers at opposite extremes of the
parameter space and observing both converging to the same area of the
posterior indicates that the underlying Markov chains have converged.
This process was used for the simulated data sets; trace plots for the
number of blocks in each case are shown in
Figure~\ref{fig:convergence-bounding}.

\section{Real data}\label{sec:unb:real-data}
The split-merge sampler is demonstrated on real networks: a network of
brain connectivity with binary edge weights in
Section~\ref{sec:unb:macaque} and a network of emails with count data
for edge weights in Section~\ref{sec:unb:enron}.

\subsection{Macaque sensory data}\label{sec:unb:macaque}
The first data set analysed concerns the brain of a macaque monkey
\citep{ngyessy2006macaque}.  Regions of the cortex were deemed
connected, or not, during a sensory task.  In total, 45 regions of the
brain were analysed as a network.

A block model was proposed to partition the regions of the brain.
This model assigns regions of the brain to the same block if their
neural activity is similar.  Since the data only provides binary
edge weights, a Bernoulli-SBM is applied.  A Beta(1,1) prior was
placed on the edge probability parameters $\theta_k$ and a DMA(1,6)
prior is placed on $(K, \bZ)$ for the block structure, thus the prior
expected number of blocks is five.  The split-merge algorithm was run
for 10,000 iterations to provide samples from the posterior
distribution of both block membership and parameter values. 1500
samples were discarded as burn-in.

\begin{figure}[ht!]
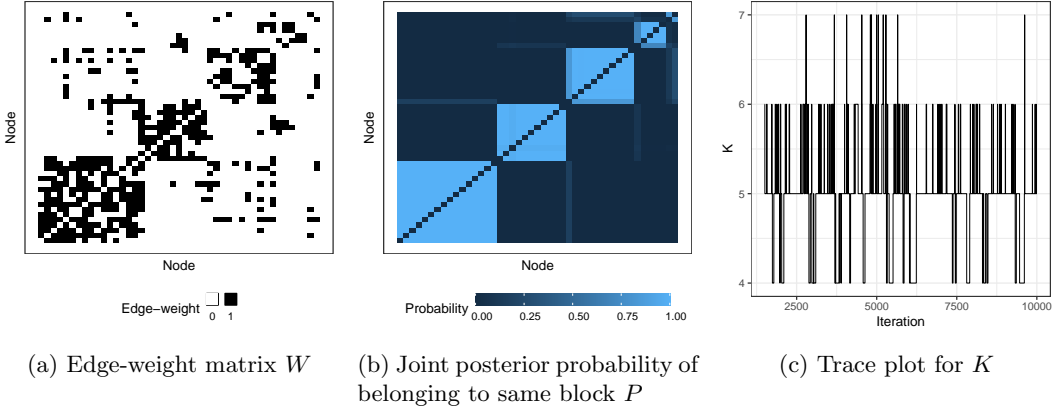

  \setlength{\subfigs}{0.3\textwidth}
  \centering
  \begin{subfigure}[t]{\subfigs}
    \centering
    \includegraphics[width=\textwidth]{{{macaque/rj_edges_10000}}}
    \caption{Edge-weight matrix $W$}
    \label{fig:macaque-edges-rj}
  \end{subfigure}%
  ~
  \begin{subfigure}[t]{\subfigs}
    \centering
    \includegraphics[width=\textwidth]{{{macaque/rj_post_pairs_10000}}}
      \caption{Joint posterior probability of belonging to same block $P$}
      \label{fig:macaque-pp-rj}
  \end{subfigure}%
  ~
  \begin{subfigure}[t]{\subfigs}
    \centering
    \includegraphics[width=\textwidth]{{{macaque/rj_num_blocks_trace_10000}}}
    \caption{Trace plot for $K$}
    \label{fig:macaque-nb-rj}
  \end{subfigure}%
  \caption{Posterior summaries for block membership in macaque brain
    network ordered by modal block assignments.}
  \label{fig:macaque-post-rj}
\end{figure}

Figure~\ref{fig:macaque-post-rj} displays posterior summaries for the
split-merge sampler.  A trace plot for the number of blocks, $K$, is
shown in Figure~\ref{fig:macaque-nb-rj}. This shows that the sampler
settles on between four and six blocks with mode five.  The joint
posterior probability matrix $P$ was calculated using
Equation~(\ref{eq:post_pairs}) and the modal block assignments were
calculated from the MCMC chain output.  Using the modal assignments,
the nodes are ordered by block label. This ordering applied to the
edge-weight matrix $W$ and $P$ are shown in
Figure~\ref{fig:macaque-edges-rj} and \ref{fig:macaque-pp-rj}
respectively.  The five blocks can be seen in
Figure~\ref{fig:macaque-pp-rj} as shown by the light blue
regions. Counting from the lower left of
Figure~\ref{fig:macaque-pp-rj}, block five consists of two nodes;
these nodes also have some probability of belonging to block three, as
indicated by the shading in the final two columns/rows. Similarly,
some uncertainty is displayed in the block membership of the first
nodes in blocks three and four. Modal parameter estimates are shown
in Table~\ref{tab:macaque-params} together with 5\% and 95\% quantiles
and the effective sample size. The parameters for smaller blocks have
wider confidence intervals; this is expected since there are fewer
edge weights governed by those parameters.  Note that parameter $\theta_5$ is
more uncertain; this is due to the block consisting of two nodes,
meaning that $\theta_5$ only governs one edge weight. The effective sample size cannot be
computed for this parameter since it is absent in many iterations when
the block has been merged with another block.

\begin{table}[ht!]
  \caption{Modal parameter estimates, 95\% posterior quantiles and
    effective sample sizes for macaque network.}
  \label{tab:macaque-params}
  \centering
  \begin{tabular}{lrrrrr}
    \hline
    Parameter & Mode & 5\% & 95\% & Effective sample size\\
    \hline
    $\theta_0$ & 0.09 & 0.08 & 0.11 & 1048\\
    $\theta_1$ & 0.70 & 0.64 & 0.75 & 553\\
    $\theta_2$ & 0.72 & 0.63 & 0.80 & 251\\
    $\theta_3$ & 0.56 & 0.43 & 0.68 & 126\\
    $\theta_4$ & 0.58 & 0.36 & 0.82 & 71\\
    $\theta_5$ & 0.70 & 0.15 & 0.99 & NA\\
    \hline
  \end{tabular}
\end{table}

\subsection{Enron emails}\label{sec:unb:enron}
The Enron corporation was declared bankrupt in 2001 and later multiple
employees were found guilty of accounting fraud.  As a result of the
trial, a corpus of emails leading up to the closure of the company was
released as a public data set \citep{klimt2004}.  Aggregate counts of
emails between any two employees are arranged into an edge-weight
matrix. Note that this network contains directed edges and self-loops
(since some emails are sent to mailing lists, to which the sender
belongs).  Two models for the edge weights were considered for this
model: (i) a Poisson with a Gamma(1,1) prior and (ii) a negative
binomial with a Gamma(1,1) prior for $r$ and a Beta(1,1) prior for
$p$.  In both cases a DMA(1,10) joint prior is placed on $K, \bZ$.  On
a first analysis, the mean number of emails sent by any one employee
is 3.7, whilst the variance is 4753, so a Poisson model seems a bad
fit \emph{a priori}.  The split-merge algorithm of
Section~\ref{sec:unb:split-merge} was applied with 10,000 iterations
and 1500 discarded as burn-in.

\begin{figure}[ht!]
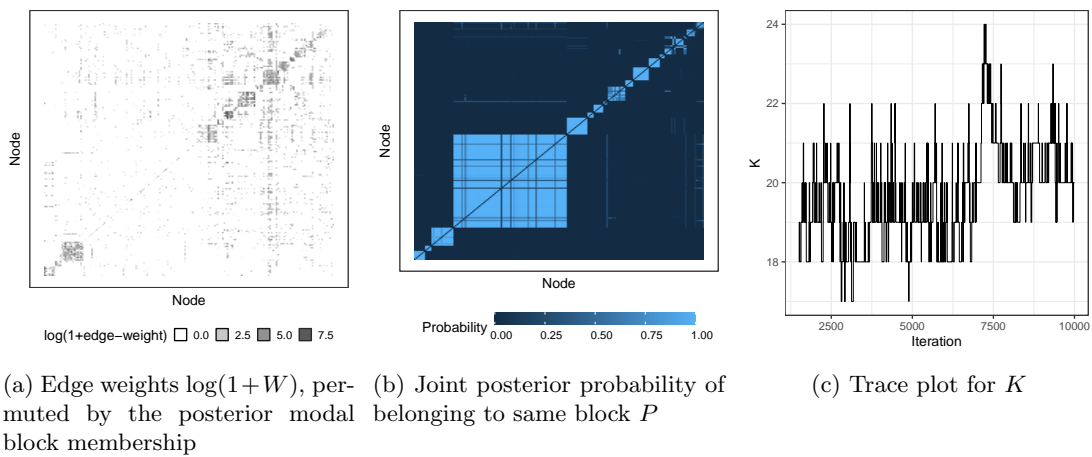

  \setlength{\subfigs}{0.31\textwidth}
  \centering
  \begin{subfigure}[t]{\subfigs}
    \centering
    \includegraphics[width=\textwidth]{{{enron/rj_pois_edges_10000}}}
    \caption{Edge weights $\log(1 + W)$, permuted by the posterior modal block membership}
    \label{fig:enron-pois-edges}
  \end{subfigure}%
  ~
  \begin{subfigure}[t]{\subfigs}
    \centering
    \includegraphics[width=\textwidth]{{{enron/rj_pois_post_pairs_10000}}}
    \caption{Joint posterior probability of belonging to same block $P$}
    \label{fig:enron-pois-pp}
  \end{subfigure}%
  ~
  \begin{subfigure}[t]{\subfigs}
    \centering
    \includegraphics[width=\textwidth]{{{enron/rj_pois_num_blocks_trace_10000}}}
    \caption{Trace plot for $K$}
    \label{fig:enron-pois-nb}
  \end{subfigure}%
  \caption{Posterior summaries for block membership in Enron network with Poisson edge-weight model (after burn-in).}
  \label{fig:enron-pois-post}
\end{figure}

\begin{figure}[ht!]
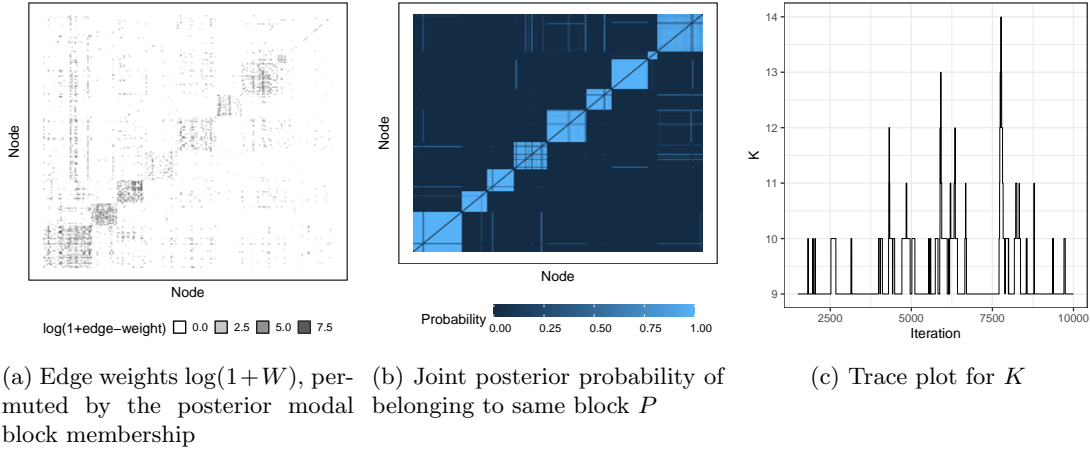

  \setlength{\subfigs}{0.31\textwidth}
  \centering
  \begin{subfigure}[t]{\subfigs}
    \centering
    \includegraphics[width=\textwidth]{{{enron/rj_nbin_edges_10000}}}
    \caption{Edge weights $\log(1 + W)$, permuted by the posterior modal block membership}
    \label{fig:enron-nbin-edges}
  \end{subfigure}%
  ~
  \begin{subfigure}[t]{\subfigs}
    \centering
    \includegraphics[width=\textwidth]{{{enron/rj_nbin_post_pairs_10000}}}
    \caption{Joint posterior probability of belonging to same block $P$}
    \label{fig:enron-nbin-pp}
  \end{subfigure}%
  ~
  \begin{subfigure}[t]{\subfigs}
    \centering
    \includegraphics[width=\textwidth]{{{enron/rj_nbin_num_blocks_trace_10000}}}
    \caption{Trace plot for $K$}
    \label{fig:enron-nbin-nb}
  \end{subfigure}%
  \caption{Posterior summaries for block membership in Enron network
    with negative binomial edge-weight model (after burn-in).}
  \label{fig:enron-nbin-post}
\end{figure}

As in Section~\ref{sec:unb:macaque}, the joint posterior probability
matrix $P$ was calculated using Equation~(\ref{eq:post_pairs}) and the
modal block assignments were calculated from the MCMC chain output.
Using the modal assignments, the nodes are ordered by block
label. This ordering applied to the log edge-weight matrix $W$ and $P$
in Figure~\ref{fig:enron-pois-edges} and
Figure~\ref{fig:enron-pois-pp} respectively.  The negative binomial
model is more flexible and is thus able to more easily detect
structure in the network compared to the Poisson model. This is exemplified
in the ordered plot of the log edge weights in Figures
\ref{fig:enron-pois-edges} and \ref{fig:enron-nbin-edges}.
Furthermore, the fit using the Poisson distribution for edge weights
finds one large group (fourth from the left in
Figure~\ref{fig:enron-pois-pp}) with a low incidence of sent
emails. This group corresponds to parameter $\lambda_4$, which has a
posterior mode of 0.19.  Under the negative binomial distribution, the
low-incidence group is much smaller, with modal parameters
$r_9 = 0.004$ and $p_9 = 0.012$ giving an expected number of emails
sent by a node in block nine as $r(1-p)/p \simeq 0.33$. The modal
parameter values for each model are given in Table~\ref{tab:enron}
together with the 5\% and 95\% quantiles.

\begin{table}[ht!]
  \caption{Parameter mode, 5\% and 95\% posterior quantiles for the
    Enron data with edge-weight model: (i) Possion($\lambda$) and (ii)
    NegativeBinomial($r, p$)
  }
  \label{tab:enron}
    \centering
    \begin{tabular}{rrrr}
      \hline
      $\theta$ & Mode & 5\% & 95\% \\
      \hline
      $r_0$ & 0.012 & 0.011 & 0.012 \\
      $r_1$ & 0.133 & 0.122 & 0.147 \\
      $r_2$ & 0.323 & 0.282 & 0.374 \\
      $r_3$ & 0.169 & 0.149 & 0.194 \\
      $r_4$ & 0.086 & 0.069 & 0.106 \\
      $r_5$ & 0.082 & 0.070 & 0.100 \\
      $r_6$ & 0.114 & 0.092 & 0.139 \\
      $r_7$ & 0.120 & 0.104 & 0.137 \\
      $r_8$ & 0.460 & 0.259 & 0.706 \\
      $r_9$ & 0.004 & 0.002 & 0.022 \\
      \hline
      $p_0$ & 0.013 & 0.012 & 0.015 \\
      $p_1$ & 0.003 & 0.002 & 0.003 \\
      $p_2$ & 0.007 & 0.006 & 0.009 \\
      $p_3$ & 0.002 & 0.002 & 0.003 \\
      $p_4$ & 0.020 & 0.014 & 0.029 \\
      $p_5$ & 0.007 & 0.005 & 0.010 \\
      $p_6$ & 0.008 & 0.005 & 0.011 \\
      $p_7$ & 0.006 & 0.005 & 0.008 \\
      $p_8$ & 0.039 & 0.019 & 0.064 \\
      $p_9$ & 0.012 & 0.001 & 0.041 \\
      \hline
      $\lambda_0$ & 1.45 & 1.39 & 1.50 \\
      $\lambda_1$ & 43.67 & 41.49 & 45.29 \\
      $\lambda_2$ & 32.43 & 30.33 & 34.70 \\
      $\lambda_3$ & 52.62 & 51.69 & 57.98 \\
      $\lambda_4$ & 0.19 & 0.15 & 0.23 \\
      $\lambda_5$ & 30.28 & 27.15 & 31.14 \\
      $\lambda_6$ & 146.85 & 142.71 & 151.71 \\
      $\lambda_7$ & 498.32 & 492.65 & 505.24 \\
      $\lambda_8$ & 29.51 & 20.73 & 174.72 \\
      $\lambda_9$ & 161.93 & 23.59 & 343.92 \\
      \hline
    \end{tabular}
\end{table}

\section{Concluding remarks}
\label{sec:unb:concluding-remarks}
This paper considered a generalisation of the stochastic block model
by allowing arbitrary edge-weight distributions and explicitly
modelling the number of blocks.  A Bayesian inference algorithm was
proposed: a split-merge reversible jump
Markov chain Monte Carlo sampler as described in Section~\ref{sec:unb:split-merge}.
Unlike previous Bayesian treatments of the stochastic block model with
an unknown number of blocks \citep{morup2012,schmidt2013,mcdaid2013},
the proposed algorithm handles edge-weight distributions without
conjugate priors.  This allows for more flexible modelling of network
data, as demonstrated in Section~\ref{sec:unb:enron} on the Enron
email network.  In this example, a negative binomial model (with both
parameters unknown) was fit to the edge weights, allowing for a higher
variance of edge weights within a block than under the Poisson
model. In the Enron data set, the negative binomial explored the
parameter space better than the Poisson model since it visited
posterior states with more structure.

The algorithm presented here is general and can be applied to the generalised
stochastic block model with any edge-weight distributions from which
samples can be taken and densities evaluated.  This can easily include
co-variate information in either the edge-weight distribution, $G$, or
the block membership distribution, $F$.

For simplicity, the models presented in Section~\ref{sec:unb:model}
assume all edges are present in the network and that each edges has a
recorded edge weight.  This assumption can be relaxed in (at least)
two ways.
Firstly, if some set of edges $\cA$ is known to be absent
from the network, then the set of edges is $\cE_A = \cE/\cA$. For
example, consider a network of electrical cables between substations.
The substations are represented by nodes, the cables by edges and the
voltage along a cable by an edge weight.
In this case, Equation~\ref{eq:restricted-sbm} remains
unchanged except the last line runs over all $ij \in \cE_A$ rather
than $\cE$. To adapt the split-merge sampler, the likelihood calculations involving
node $i$ iterate over all nodes $j \in \cE_A/\set{i}$ instead of all $i
\neq j$.
In the second case, the edge exists in the model but the edge weight
is not recorded in the data set; this is a missing data problem. Two
approaches are possible: either the edge weight was not
recorded, or the edge does not exist.  In the first case, one could
use a data augmentation scheme within the split-merge sampler to infer
the state of missing edge weights. In the second case, a sparsity
parameter as in \citet{matias2016} could be inferred within the
GSBM framework. This treats edge weights as a mixture of the density
$G$ and a Dirac mass at zero representing the non-existence of an
edge.

\section*{Acknowledgements}
The author would like to thank the referees as well as Brendan Murphy,
Simon Lunagomez and Peter Neal for helpful comments.
Funding: This work was supported by the Engineering and Physical Sciences Research Council (EPSRC) [EP/H023151/1 and EP/P033075/1].

\appendix
\section{Acceptance probability calculations}
\label{app:accept-prob-calc}

Since a merge move is the inverse of a split move, $A_{merge} = 1/A_{split}$, hence only $A_{split}$ is derived.
The acceptance probability can be split into the following parts: posterior density ratio, proposal density ratio, ratio of densities of auxiliary variables, and the Jacobian; as such $A_{split}$ has the general form:
\begin{equation}
  \label{eq:rj-a-split-full}
  \begin{aligned}
    A_{split} &= \frac{\post{\kappa+1, \bz', \btheta' | E}} {\post{\kappa, \bz, \btheta | E}}
    \frac{\prop{\kappa, \bz, \btheta | \kappa+1, \bz', \btheta'}} {\prop{\kappa+1, \bz', \btheta' | \kappa, \bz, \btheta}}
    \frac{\prop{\lambda}} {\prop{u',\lambda'}} J_{split}\\
    &=  \frac{\post{\kappa+1, \bz', \btheta' | E}} {\post{\kappa, \bz, \btheta | E}}
    \frac{\prop{merge | \kappa+1}} {\prop{split | \kappa}}
    \frac{\prop{k',l'}} {\prop{k}}
    \frac{\prop{\lambda}} {\prop{\lambda', u'}}
    \frac{1}{\prop{\bz'|\btheta'}}
    J_{split}\\
  \end{aligned}
\end{equation}
where $\prop{split | \kappa}$ and $\prop{merge | \kappa}$ are the
probabilities of proposing a split or merge move given that the
current state of the sampler contains $\kappa$ blocks.  These are
chosen as 1/2 where possible.  That is $\prop{split | \kappa = 1} = 1$
and $\prop{merge | \kappa=1} = 0$ since merging is impossible when
there is only one block.
Note that in the examples: $\lambda,\lambda'\simiid \Unif{0,1}$, $u' \sim \Normal{0,1}$, $k'$ and
$k,l$ are sampled at random amongst the set of available blocks.

Finally, $J_{split}$ is the Jacobian of the split proposal given in Equation~(\ref{eq:rj-j-split}) and $p$ is the dimensionality of each $\btheta_k$.
\begin{equation}
  \label{eq:rj-j-split}
  J_{split} =
  \begin{vmatrix}
    \partfrac{\btheta'_{k'}}{\btheta_k} & \partfrac{\btheta'_{l'}}{\btheta_k} \\[2ex]
    \partfrac{\btheta'_{k'}}{u'}        & \partfrac{\btheta'_{l'}}{u'}
  \end{vmatrix} = \left| \frac{\grad{m}{\btheta'_{k'}}\grad{m}{\btheta'_{l'}}} {\grad{m}{\btheta_k} (2\lambda(1-\lambda))^p} \right|
\end{equation}

Therefore, in the examples, where specific choices for $u',\lambda',\lambda$ and $\prop{merge}, \prop{split}$ have been made, the acceptance probabilities reduce to:
\begin{align*}
  A_{split} =& \frac{\post{\kappa+1, \bz', \btheta' | E}} {\post{\kappa, \bz, \btheta | E}}
               \frac{1}{1 + \II{\kappa=1}}
               \frac{2}{\kappa+1}\\
             &\qquad\times\frac{1}{\fp{\phi}{u' | 0, \sigma^2}}
               \frac{1}{\prop{\bz'| \btheta'}}
               \left| \frac{\grad{m}{\btheta'_{k'}}\grad{m}{\btheta'_{l'}}} {\grad{m}{\btheta_k} (2\lambda(1-\lambda))^p} \right|\\
  A_{merge} &= \frac{\post{\kappa-1, \bz', \btheta' | E}} {\post{\kappa, \bz, \btheta | E}}
              \left(1 + \II{\kappa=2}\right)
              \frac{\kappa}{2}\\
             &\qquad\times \fp{\phi}{u | 0, \sigma^2}
               \prop{\bz |\btheta}
               \left| \frac{\grad{m}{\btheta'_{k'}} (2\lambda(1-\lambda))^p} {\grad{m}{\btheta_{k}}\grad{m}{\btheta_{l}}} \right|
\end{align*}

\section{Post-hoc matching}
\label{app:matching}
The GSBM is invariant to relabelling of the nodes --
Equation~\ref{eq:general-sbm} gives the same posterior value if the node
labels are permuted. This causes a problems when comparing the output
of the MCMC against some known parameter values in
Section~\ref{sec:unb:application}, since the estimated block labels need to
match the truth for a reasonable comparison.

Let $Z^{\text{true}}$ be a set of true block labels. We match the MCMC
output labels to the true labels by matching the modal assignment
vector $Z^{\text{mode}}$ to $Z^{\text{true}}$, where
$$
Z_i^{\text{mode}} = \arg\max_{k}\sum_{\cS} \II{Z_{is}= k},
$$
gives the most-often used block label for node $i$ during the MCMC
iterations in $\cS$.

Given $Z^{\text{true}}$ and $Z^{\text{mode}}$, a contingency
table $n$ is formed via:
$$
n_{ck} = \sum_{i}\II{(Z_i^{\text{mode}} = c) \& (Z_i^{\text{true}} = k)}.
$$
Thus entry $c,k$ in the table is the number of nodes assigned to block
$c$ under the mode and block $k$ under the truth.

Let $\pi$ be a permutation with $\pi_c = \arg\max_k n_{ck}$.
We relabel the MCMC output for each $i=1 \to N$ and $s \in \cS$ via
$Z_{is} = c \mapsto Z_{is} = \pi_c$ and $\theta_c \mapsto \theta_{\pi_c}$.
Under this relabelling the modal and true labels match so comparisons
between parameters can be made.

\bibliographystyle{elsarticle-harv}
\bibliography{./paper/refs}

\begin{thebibliography}{40}
\expandafter\ifx\csname natexlab\endcsname\relax\def\natexlab#1{#1}\fi
\expandafter\ifx\csname url\endcsname\relax
  \def\url#1{\texttt{#1}}\fi
\expandafter\ifx\csname urlprefix\endcsname\relax\def\urlprefix{URL }\fi

\bibitem[{Airoldi et~al.(2008)Airoldi, Blei, Fienberg, and Xing}]{airoldi2008}
Airoldi, E.~M., Blei, D.~M., Fienberg, S.~E., Xing, E.~P., 2008. Mixed
  membership stochastic blockmodels. Journal of Machine Learning Research
  9~(Sep), 1981--2014.

\bibitem[{Ambroise and Matias(2012)}]{ambroise2012}
Ambroise, C., Matias, C., 2012. New consistent and asymptotically normal
  parameter estimates for random-graph mixture models. Journal of the Royal
  Statistical Society: Series B (Statistical Methodology) 74~(1), 3--35.

\bibitem[{Chen and Lei(2016)}]{chen2016}
Chen, K., Lei, J., 2016. Network cross-validation for determining the number of
  communities in network data. Journal of the American Statistical Association,
  1--11.
\newline\urlprefix\url{https://doi.org/10.1080/01621459.2016.1246365}

\bibitem[{Copic et~al.(2009)Copic, Jackson, and Kirman}]{copic2009}
Copic, J., Jackson, M.~O., Kirman, A., 2009. Identifying community structures
  from network data via maximum likelihood methods. The BE Journal of
  Theoretical Economics 9~(1).

\bibitem[{Daudin et~al.(2008)Daudin, Picard, and Robin}]{daudin2008}
Daudin, J.-J., Picard, F., Robin, S., 2008. A mixture model for random graphs.
  Statistics and Computing 18~(2), 173--183.
\newline\urlprefix\url{https://doi.org/10.1007/s11222-007-9046-7}

\bibitem[{Fienberg et~al.(1985)Fienberg, Meyer, and Wasserman}]{fienberg1985}
Fienberg, S.~E., Meyer, M.~M., Wasserman, S.~S., 1985. Statistical analysis of
  multiple sociometric relations. Journal of the american Statistical
  association 80~(389), 51--67.

\bibitem[{Frank and Harary(1982)}]{frank1982}
Frank, O., Harary, F., 1982. Cluster inference by using transitivity indices in
  empirical graphs. Journal of the American Statistical Association 77~(380),
  835--840.

\bibitem[{Gelman and Rubin(1992)}]{gelman1992}
Gelman, A., Rubin, D.~B., 1992. Inference from iterative simulation using
  multiple sequences. Statist. Sci. 7~(4), 457--472.
\newline\urlprefix\url{https://doi.org/10.1214/ss/1177011136}

\bibitem[{Geng et~al.(2019)Geng, Bhattacharya, and Pati}]{Geng2019}
Geng, J., Bhattacharya, A., Pati, D., 2019. Probabilistic community detection
  with unknown number of communities. Journal of the American Statistical
  Association 114~(526), 893--905.

\bibitem[{Gershman and Blei(2012)}]{gershman2012}
Gershman, S.~J., Blei, D.~M., 2012. A tutorial on {B}ayesian nonparametric
  models. Journal of Mathematical Psychology 56~(1), 1 -- 12.
\newline\urlprefix\url{https://doi.org/10.1016/j.jmp.2011.08.004}

\bibitem[{Green(1995)}]{green1995}
Green, P.~J., 1995. Reversible jump {M}arkov chain {M}onte {C}arlo computation
  and {B}ayesian model determination. Biometrika 82~(4), 711--732.

\bibitem[{Green and Richardson(2001)}]{green2001}
Green, P.~J., Richardson, S., 2001. Modelling heterogeneity with and without
  the {D}irichlet process. Scandinavian Journal of Statistics 28~(2), 355--375.
\newline\urlprefix\url{http://dx.doi.org/10.1111/1467-9469.00242}

\bibitem[{Hoff et~al.(2002)Hoff, Raftery, and Handcock}]{hoff2002}
Hoff, P.~D., Raftery, A.~E., Handcock, M.~S., 2002. Latent space approaches to
  social network analysis. Journal of the American Statistical Association
  97~(460), 1090--1098.
\newline\urlprefix\url{https://doi.org/10.1198/016214502388618906}

\bibitem[{Holland et~al.(1983)Holland, Laskey, and Leinhardt}]{holland1983}
Holland, P.~W., Laskey, K.~B., Leinhardt, S., 1983. Stochastic blockmodels:
  First steps. Social networks 5~(2), 109--137.

\bibitem[{Jiang et~al.(2009)Jiang, Zhang, and Sun}]{jiang2009}
Jiang, Q., Zhang, Y., Sun, M., 2009. Community detection on weighted networks:
  A variational {B}ayesian method. In: Asian Conference on Machine Learning.
  Springer, pp. 176--190.

\bibitem[{Karrer and Newman(2011)}]{karrer2011}
Karrer, B., Newman, M.~E., 2011. Stochastic blockmodels and community structure
  in networks. Physical Review E 83~(1), 016107.

\bibitem[{Kemp et~al.(2006)Kemp, Tenenbaum, Griffiths, Yamada, and
  Ueda}]{kemp2006}
Kemp, C., Tenenbaum, J.~B., Griffiths, T.~L., Yamada, T., Ueda, N., 2006.
  Learning systems of concepts with an infinite relational model. In: AAAI.
  Vol.~3. p.~5.

\bibitem[{Klimt and Yang(2004)}]{klimt2004}
Klimt, B., Yang, Y., 2004. Machine Learning: ECML 2004: 15th European
  Conference on Machine Learning, Pisa, Italy, September 20-24, 2004.
  Proceedings. Springer Berlin Heidelberg, Berlin, Heidelberg, Ch. The Enron
  Corpus: A New Dataset for Email Classification Research, pp. 217--226.

\bibitem[{Latouche et~al.(2012)Latouche, Birmele, and Ambroise}]{latouche2012}
Latouche, P., Birmele, E., Ambroise, C., 2012. Variational {B}ayesian inference
  and complexity control for stochastic block models. Statistical Modelling
  12~(1), 93--115.

\bibitem[{Lei(2016)}]{lei2016}
Lei, J., 2016. A goodness-of-fit test for stochastic block models. The Annals
  of Statistics 44~(1), 401--424.
\newline\urlprefix\url{https://doi.org/10.1214/15-aos1370}

\bibitem[{Ludkin(2020)}]{CRANSBMSplitMerge}
Ludkin, M., 2020. SBMSplitMerge: Inference for a Generalised {SBM} with a Split
  Merge Sampler. R package version 1.1.1.
\newline\urlprefix\url{https://cran.r-project.org/package=SBMSplitMerge}

\bibitem[{Ludkin et~al.(2018)Ludkin, Eckley, and Neal}]{ludkin2017}
Ludkin, M., Eckley, I., Neal, P., 2018. Dynamic stochastic block models:
  parameter estimation and detection of changes in community structure.
  Statistics and Computing.
\newline\urlprefix\url{https://doi.org/10.1007/s11222-017-9788-9}

\bibitem[{Mariadassou et~al.(2010)Mariadassou, Robin, and
  Vacher}]{mariadassou2010}
Mariadassou, M., Robin, S., Vacher, C., 2010. Uncovering latent structure in
  valued graphs: a variational approach. The Annals of Applied Statistics,
  715--742.

\bibitem[{Matias and Miele(2017)}]{matias2016}
Matias, C., Miele, V., 2017. Statistical clustering of temporal networks
  through a dynamic stochastic block model. Journal of the Royal Statistical
  Society: Series B (Statistical Methodology) 79~(4), 1119--1141.

\bibitem[{Matias and Robin(2014)}]{matias2014}
Matias, C., Robin, S., 2014. Modeling heterogeneity in random graphs through
  latent space models: a selective review. ESAIM: Proc. 47, 55--74.
\newline\urlprefix\url{https://doi.org/10.1051/proc/201447004}

\bibitem[{Mc{D}aid et~al.(2013)Mc{D}aid, Murphy, Friel, and
  Hurley}]{mcdaid2013}
Mc{D}aid, A.~F., Murphy, T.~B., Friel, N., Hurley, N.~J., 2013. Improved
  {B}ayesian inference for the stochastic block model with application to large
  networks. Computational Statistics \& Data Analysis 60, 12--31.
\newline\urlprefix\url{http://dx.doi.org/10.1016/j.csda.2012.10.021}

\bibitem[{Miller and Harrison(2018)}]{Miller2018}
Miller, J.~W., Harrison, M.~T., 2018. Mixture models with a prior on the number
  of components. Journal of the American Statistical Association 113~(521),
  340--356.

\bibitem[{M{\o}rup and Schmidt(2012)}]{morup2012}
M{\o}rup, M., Schmidt, M.~N., 2012. {B}ayesian community detection. Neural
  computation 24~(9), 2434--2456.

\bibitem[{M{\o}rup and Schmidt(2013)}]{schmidt2013}
M{\o}rup, M., Schmidt, M.~N., 2013. Nonparametric {B}ayesian modeling of
  complex networks: an introduction. {IEEE} Signal Processing Magazine 30~(3),
  110--128.

\bibitem[{M{\o}rup et~al.(2011)M{\o}rup, Schmidt, and Hansen}]{morup2011}
M{\o}rup, M., Schmidt, M.~N., Hansen, L.~K., 2011. Infinite multiple membership
  relational modeling for complex networks. In: Machine Learning for Signal
  Processing (MLSP), 2011 IEEE International Workshop on. IEEE, pp. 1--6.

\bibitem[{N{\'{e}}gyessy et~al.(2006)N{\'{e}}gyessy, Nepusz, Kocsis, and
  Bazs{\'{o}}}]{ngyessy2006macaque}
N{\'{e}}gyessy, L., Nepusz, T., Kocsis, L., Bazs{\'{o}}, F., 2006. Prediction
  of the main cortical areas and connections involved in the tactile function
  of the visual cortex by network analysis. European Journal of Neuroscience
  23~(7), 1919--1930.

\bibitem[{Nobile and Fearnside(2007)}]{nobile2007}
Nobile, A., Fearnside, A.~T., 2007. {B}ayesian finite mixtures with an unknown
  number of components: The allocation sampler. Statistics and Computing
  17~(2), 147--162.
\newline\urlprefix\url{https://doi.org/10.1007/s11222-006-9014-7}

\bibitem[{Nowicki and Snijders(2001)}]{nowicki2001}
Nowicki, K., Snijders, T. A.~B., 2001. Estimation and prediction for stochastic
  blockstructures. Journal of the American Statistical Association 96~(455),
  1077--1087.
\newline\urlprefix\url{https://doi.org/10.1198/016214501753208735}

\bibitem[{Peixoto(2013)}]{peixoto2013}
Peixoto, T.~P., 2013. Parsimonious module inference in large networks. Physical
  Review Letters 110~(14).
\newline\urlprefix\url{https://doi.org/10.1103/physrevlett.110.148701}

\bibitem[{Salda{\~{n}}a et~al.(2017)Salda{\~{n}}a, Yu, and Feng}]{saldaa2017}
Salda{\~{n}}a, D.~F., Yu, Y., Feng, Y., 2017. How many communities are there?
  Journal of Computational and Graphical Statistics 26~(1), 171--181.
\newline\urlprefix\url{https://doi.org/10.1080/10618600.2015.1096790}

\bibitem[{Snijders and Nowicki(1997)}]{snijders1997}
Snijders, T.~A., Nowicki, K., 1997. Estimation and prediction for stochastic
  blockmodels for graphs with latent block structure. Journal of classification
  14~(1), 75--100.

\bibitem[{Snijders et~al.(2006)Snijders, Pattison, Robins, and
  Handcock}]{snijders2006}
Snijders, T. A.~B., Pattison, P.~E., Robins, G.~L., Handcock, M.~S., 2006. New
  specifications for exponential random graph models. Sociological Methodology
  36~(1), 99--153.

\bibitem[{Wang et~al.(2017)Wang, Bickel, et~al.}]{wang2017}
Wang, Y.~R., Bickel, P.~J., et~al., 2017. Likelihood-based model selection for
  stochastic block models. The Annals of Statistics 45~(2), 500--528.

\bibitem[{Wasserman and Anderson(1987)}]{wasserman1987}
Wasserman, S., Anderson, C., 1987. Stochastic a posteriori blockmodels:
  Construction and assessment. Social Networks 9~(1), 1--36.
\newline\urlprefix\url{https://doi.org/10.1016/0378-8733(87)90015-3}

\bibitem[{Xin et~al.(2017)Xin, Zhu, and Chipman}]{xin2017}
Xin, L., Zhu, M., Chipman, H., 2017. A continuous-time stochastic block model
  for basketball networks. The Annals of Applied Statistics 11~(2), 553--597.

\end{thebibliography}

\end{document}